\documentclass[preprint]{aastex}
\usepackage{longtable,lscape}
\usepackage{natbib}
\usepackage[dvips]{color}
\newcommand{\HII}{H\,{\sc ii}}
\newcommand{\HI}{H\,{\sc i}}
\newcommand{\Ha}{H$\alpha$}

\shortauthors{Zhao, Gu, \& Gao}
\shorttitle{The Stellar Population and star formation properties of BCDs}

\begin{document}
\title{The Stellar Population and Star Formation Properties of Blue Compact Dwarf Galaxies}
\author{Yinghe Zhao\altaffilmark{1, 2}, Qiusheng Gu\altaffilmark{2, 3, 4}, Yu Gao\altaffilmark{1, 4}}
\altaffiltext{1}{Purple Mountain Observatory, Chinese Academy of Sciences (CAS), Nanjing 210008, China; yhzhao, yugao@pmo.ac.cn}
\altaffiltext{2}{Key Laboratory of Modern Astronomy and Astrophysics (Nanjing University), Ministry of Education, Nanjing 210093, China}
\altaffiltext{3}{Department of Astronomy, Nanjing University, Nanjing 210093, China; qsgu@nju.edu.cn}
\altaffiltext{4}{Author to whom correspondence should be addressed}

\begin{abstract}
We study the stellar populations, star formation histories and star formation properties for a sample of blue compact dwarf galaxies (BCDs) selected by cross-correlating the Gil de Paz et al. (2003) sample with the Sloan Digital Sky Survey Data Release 6 (SDSS DR6). The sample includes 31 BCDs, which span a large range in galactic parameters. Using a stellar population synthesis method, we derive the stellar populations and reconstruct the star formation histories for these BCDs. Our studies confirm that BCDs are not young systems experiencing their first star formation but old systems undergoing a starburst activity. The stellar mass-weighted ages can be as old as 10 Gyr while the luminosity-weighted ages might be up to $\sim 3$ orders of magnitude younger ($\sim 10$ Myr) for most galaxies.

Based on multi-wavelength data, we also study the integrated star formation properties. The SFR for our sample galaxies spans nearly 3 orders of magnitude, from a few $10^{-3}$ to $\sim1\,M_\odot$ yr$^{-1}$, with the median value of $\sim 0.1\, M_\odot$ yr$^{-1}$. We find that about 90\% BCDs in our sample have their birthrate parameter (the ratio of the current SFR to the averaged past SFR) $b>2-3$. We further discuss correlations of the current SFR with the integrated galactic stellar mass and explore the connection between SFR and metallicity.
\end{abstract}
\keywords{galaxies: dwarf -- galaxies: starburst -- galaxies: stellar content}

\section{Introduction}
Dwarf galaxies play an important role in our understanding of the formation and evolution of galaxies. In the hierarchical model of galaxy formation, they are proposed to be the building-blocks from which larger systems have been created by merging (Kauffmann et al. 1993). However, these building-block galaxies are too small and faint to be studied at high redshifts. Therefore, of special interest are the studies aimed at the search for some local example galaxies which might have much in common with the galaxies at high redshifts. 

Blue compact dwarf galaxies (BCDs) are such local example galaxies, whose characteristics are believed to have been common among unevolved low-mass galaxies at intermediate to high redshift. BCDs are small, gas-rich (H\,{\sc i} mass fraction typically higher than 30\%; e.g. Thuan \& Martin 1981; Salzer et al. 2002) and metal-poor ($1/50 \leq Z \leq 1/3\ Z_\odot$; Hunter \& Hoffman 1999) extragalactic objects. They have dramatically different properties compared to normal dwarf galaxies (Zwicky 1966; Gil de Paz et al. 2003, hereafter G03), and are spectroscopically characterized by a faint, blue optical continuum accompanied, in most cases, by strong narrow emission lines, due to the intense star formation activity in one or several star-forming regions (Cair\'os et al. 2001). Stars are formed at high rates in BCDs (Fanelli et al. 1988), exhausting their gas content with a timescale of $10^9$ yr (e.g. Lee et al. 2002), which is much shorter than the age of the Universe. This fact, combined with the low metal abundances, led Sargent \& Searle (1970) to suggest that BCDs are primeval galaxies undergoing star formation for the first time.

The question as to whether BCDs are truly young galaxies or they are old galaxies exhibiting a strong starburst has been widely discussed for many years (e.g. Sargent \& Searle 1970; Searle et al. 1973; Schulte-Ladbeck et al. 2000, 2001a; Izotov \& Thuan 1999, 2004; Aloisi et al. 2005). Most BCDs are known to have a red low surface brightness background of presumably older stars (e.g., Loose \& Thuan 1986; Gil de Paz \& Madore 2005; Caon et al. 2005). Moreover, some individual red giant stars (RGBs) in the nearest BCDs can be resolved (e.g., Lynds et al. 1998; Drozdovsky et al. 2001; Corbin et al. 2008). This evidence ruled out the possibility that the majority of BCDs began to form stars within the last billion years. 

An alternative view proposed by Searle et al. (1973) is that BCDs are chemically primitive objects which experience an episodic star formation history (SFH). The star formation in BCDs occurs in intense bursts which are interleaved by long quiescent periods. This traditional picture also seems to be problematic. Several nearby BCDs with resolved stars do not show evidence of long gaps ($>1$ Gyr) in their recent star formation history (Schulte-Ladbeck et al. 2001a; Crone et al. 2002; Annibali et al. 2003; McQuinn et al. 2009). S\'anchez Almeida et al. (2008) identified a large sample of quiescent counterparts to BCDs (QBCD) in SDSS Data Release 6 (DR6; Adelman-McCarthy et al. 2008), which might support this recursive BCD phase scenario. They also argued that the quiescent phase could last 30 times longer than the starburst phase, and thus the quiescent period of BCDs might be several hundred Myrs under the assumption of a single 10 Myr long starburst per BCD phase. However, S\'anchez Almeida et al. (2008) showed that only $\sim 15\%$ of their QBCD candidates exhibit H$\alpha$ absorptions and therefore are lack of star-forming activities. Combining the results described above with the fact that very few late-type and H\,{\sc i} selected galaxies have no observable H$\alpha$ emissions (Meurer et al. 2006; James et al. 2008; Kennicutt et al. 2008), the SFR of the quiescent phase is probably not  zero, but rather may be characterized by a continuous, low level of activity as discussed in Lee et al. (2009).

The detailed studies of SFHs of BCDs are rather difficult. For the most nearby systems, where individual stars can be resolved, the Color-Magnitude Diagrams (CMDs) synthesis method could be used to reconstruct the SFHs (e.g. I Zw 18, Aloisi et al. 1999, 2007; Mrk 178, Schulte-Ladbeck et al. 2000; I Zw 36, Schulte-Ladbeck et al. 2001a; NGC 1705, Annibali et al. 2003). However, this method can not be applied to more distant objects, which are the great majority of BCDs. The only way to study their stellar populations is to compare their integrated properties with the predictions of evolutionary synthesis models. There have been several works using (evolutionary) synthesis models to study the observed spectral energy distribution of BCDs (e.g. Mas-Hess \& Kunth 1999; Guseva et al. 2001; Corbin et al. 2006). However, these studies only include one or a few galaxies. In this paper, we present a detailed study on the stellar populations of a BCD sample containing 31 galaxies, using a Simple Stellar Population (SSP) synthesis method. This is capable of yielding the various stellar components, the internal reddening (both stellar continuum and nebular line emission) and the SFH. The pure emission-line spectra allow us to measure the fluxes of emission lines more accurately. We can also study the star formation properties of these 31 BCDs by the using the continuum-subtracted, narrow band \Ha\ imaging data (G03) combined with the results presented in this work.

The paper is organized as follows: Section 2 describes the BCD sample and our data reductions. Our results and analysis are given in Section 3. We discuss the uncertainties of the stellar population synthesis and the aperture effects of the fiber spectra in Section 4, and summarize our results in the last section. Where required we adopt a Hubble constant of $H_0 = 70\ {\rm km s}^{-1}$ Mpc$^{-1}$, $\Omega_{\rm M} = 0.3$ and $\Omega_\Lambda=0.7$.

\section{Sample, Data and Method}
\subsection{Sample and Data}
There have been a number of criteria for selecting BCDs in literature. These criteria are commonly based on the galaxy's morphological properties and its luminosity (Zwicky \& Zwicky 1971; Thuan \& Martin 1981), although some definitions are based on their spectroscopic properties (Gallego et al. 1996). BCDs classified by these different criteria may be confused with each other, e.g. a spectroscopically classified BCD may not be a BCD based on its morphological or luminosity properties (see G03).
 
Using a unified concept of BCD galaxy, G03 compiled a large but not complete sample of BCDs (including 105 members with heliocentric recession velocities less than 4000 km s$^{-1}$) from several exploratory studies (Haro 1956; Zwicky \& Zwicky 1971; Smith et al. 1976; Kunth et al. 1981; MacAlpine \& Williams 1981; Markarian et al. 1981, 1986; Zamorano et al. 1994, 1996; Alonso et al. 1999; Ugryumov et al. 1999), in which the galaxies were selected or discovered by (1) the ultraviolet excess, or (2) the presence of emission lines, or (3) the compactness and blue colors. The galaxies in the G03 sample were selected by putting forward a new set of quantitative classification criteria, namely a combination of the galaxy's color, morphology and luminosity. As presented in G03, these criteria are
\begin{center}
$\mu_{B,\rm{peak}}-\mu_{R,\rm{peak}} < 1$,\\
$\mu_{B,\rm{peak}} < 22$ mag arcsec$^{-2}$, \\
$M_K > -21$ mag.
\end{center}
where $\mu_{B,\rm{peak}}$ and $\mu_{R,\rm{peak}}$ are the peak surface brightness in {\it B}- and {\it R}-band, respectively. 

By cross-correlating the BCD sample in G03 with the SDSS DR6, we obtained the optical spectra for 31 BCDs. In Table 1 we list some basic parameters for these 31 galaxies. The distances are all adopted from G03. Except for three galaxies whose distances have been determined by measuring the magnitude of the tip of the RGB (Mrk 178, Schulte-Ladbeck et al. 2000) and by using the period-luminosity relation of Cepheids (Haro 8 and VCC 0130, Macri et al. 1999), the rest were computed using the galactic standard of rest velocity of the galaxies. The gas phase oxygen abundances, both derived with direct and empirical methods, are taken from Zhao et al. (2010; hereafter ZGG10; see also the references therein). The direct method is to use electron temperature-sensitive lines (the so-called $T_e$-method), whereas the empirical method (e.g. N2-method; Denicol\'o et al. 2002) measures the oxygen abundance through empirical relation between emission line ratio(s) and metallicity. As pointed out in Denicol\'o et al. (2002) and discussed in ZGG10 (also see Pettini \& Pagel 2004 and Salzer et al. 2005), the metallicity measured with the N2-method is less precise than with the $T_e$-method, and has an accuracy of  $\sim0.2$ dex. 

%\begin{center}
\begin{deluxetable}{lccc}
\tabletypesize{\scriptsize}
\tablewidth{0pt}
\tablecaption{Basic properties of the BCD sample}
\tablehead{
\colhead{}&\colhead{Distance} &\multicolumn{2}{c}{$12+\log\,({\rm O/H})$\tablenotemark{a}}\\
\cline{3-4}
\colhead{Galaxy}&\colhead{(Mpc)}&\colhead{$T_e$}&\colhead{N2}}
\startdata
HS 0822+3542\dotfill&10.1&   $7.42\pm0.03$&7.47\\
HS 1400+3927\dotfill&21.1&   $8.11\pm0.06$&8.08\\
HS 1440+4302\dotfill&38.0&    $8.09\pm0.03$&8.08\\
HS 1609+4827\dotfill&42.6&    $8.14\pm0.14$&8.37\\
Haro 2\dotfill&21.6&$8.38\pm0.03$&8.64\\
Haro 3\dotfill&14.4&  $8.46\pm0.10$&8.39\\
Haro 8\dotfill&16.0& $8.35\pm0.04$&8.16\\
I Zw 123\dotfill&11.7&$8.07\pm0.02$&7.99\\
II Zw 71\dotfill&18.7&  $8.24\pm0.17$&8.42\\
Mrk 1313\dotfill&31.8 &$8.22\pm0.10$&8.05\\
Mrk 1416\dotfill&33.8&$7.84\pm0.02$&7.94\\
Mrk 1418\dotfill&11.4 &\nodata&8.48\\
Mrk 1423\dotfill&20.4 &\nodata&8.52\\
Mrk 1450\dotfill&14.7&$7.96\pm0.02$&7.93\\
Mrk 1480\dotfill&27.2&$8.04\pm0.05$&8.03\\
Mrk 1481\dotfill&27.3&\nodata&8.11\\
Mrk 178\dotfill& 4.2&$7.92\pm0.02$&7.88\\
Mrk 409\dotfill&21.3 &\nodata&8.80\\
Mrk 475\dotfill& 9.7&$7.93\pm0.02$&7.89\\
Mrk 67\dotfill&14.3& $8.08\pm0.08$&7.90\\
SBS 1054+504\dotfill&20.1& $8.26\pm0.10$&8.34\\
SBS 1147+520\dotfill&18.9& \nodata&8.19\\
SBS 1428+457\dotfill&35.3& $8.42\pm0.05$&8.18\\
SBS 1533+574\dotfill&49.6&$8.07\pm0.07$&8.11\\
UGCA 184\dotfill&23.0&$8.03\pm0.01$&7.89\\
UM 323\dotfill&28.3& $7.96\pm0.04$&8.00\\
UM 439\dotfill&14.0& $8.08\pm0.03$&7.93\\
UM 452\dotfill&19.5&$8.27\pm0.20$&8.40\\
UM 456A\dotfill&24.5& \nodata&8.21\\
UM 491\dotfill&27.6   &\nodata&8.31\\
VCC 0130\dotfill&16.0 &\nodata&8.28\\
\enddata
\tablenotetext{a}{Oxygen abundances derived with the $T_e$-method and N2-method, respectively, see ZGG10 for more details.}
\label{table1}
\end{deluxetable}

Therefore, our sample is not complete in any sense, which is common in all BCD studies in the literature, as it is difficult to achieve a complete sample for faint galaxies (however, see recent efforts to construct complete sample of dwarf galaxies from Blanton et al. (2005), Kennicutt et al. (2008) and Lee et al. (2009)). Fortunately, our sample is derived from the largest BCD sample to date and spans a large range in galactic parameters including integrated luminosity ($M_B$ of $-12.0$ to $-18.5$), color ($B-R$ of $0.0-1.5$), central surface brightness and size ($\mu_{R,\,0}$ of $18.5-23.5$ mag arcsec$^{-2}$ and effective radius at $R$-band of $0.09-2.2$ kpc, see Gil de Paz \& Madore 2005) of the underlying stellar population (USP), current star formation activity ($0.03-1.6$ $M_\odot$ yr$^{-1}$ kpc$^{-2}$), oxygen abundance ($12+\log ({\rm{O/H}})$ of $\sim1/20-1/2\ Z_\odot$), and relative gas content ($0.1-3.3\ M_\odot/L_{B,\,\odot}$, for 24 galaxies that have \HI\ observations in the literature, see Zhao et al. (2010, in prep. and references therein). Because of the complexity and diversity of galaxies, we need  a large sample to probe the full range in galactic parameters in the future work.

In order to check whether this subsample is representative of the total sample of G03 with respect to the integrated ($B-R$) color and the absolute $B$ magnitude ($M_B$), we plot the ($B-R$) and $M_B$ distributions for the G03 sample and our sample using blank and hatched histograms, respectively, in Figure 1. We can see that our sample has similar distributions of ($B-R$) and $M_B$ to the G03 sample, except that our sample contains a smaller fraction of galaxies with ($B-R$) bluer than 0.5.
%The Kolmogorov-Smirnov (K-S) test shows that these two populations are drawn from the same parent population with a possibility of 0.77 and 1.0, for the ($B-R$) and $M_B$, respectively. 
The $B$- and $R$-band photometric data are directly taken from G03, where the foreground Galactic extinctions have been corrected using $A_B$ values determined following Schlegel et al.  (1998) and the Galactic extinction law of Cardelli et al. (1989).

\begin{figure}[pthb]
\centering
{\includegraphics[bb=66 1 420 360]{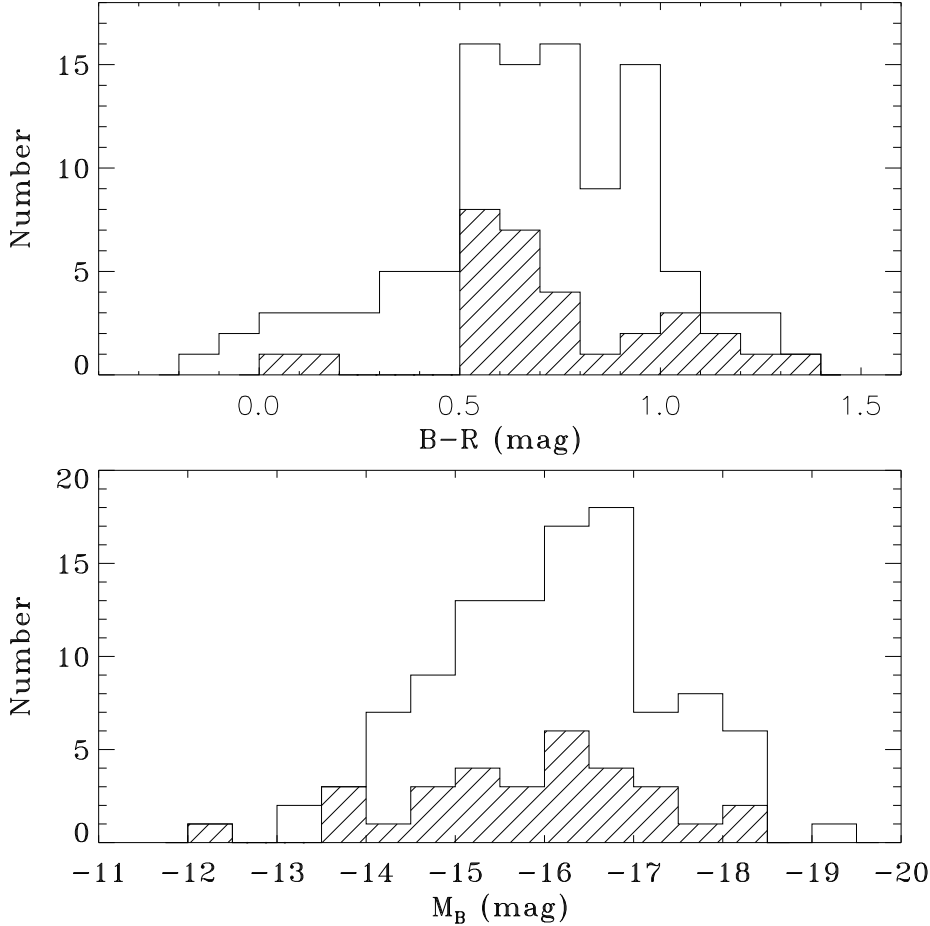}}
\caption{ Number distribution of the BCDs in $(B-R)$ color (upper panel) and absolute $B$ magnitude (bottom panel). {\it Blank histogram}: G03 sample; {\it Hatched histogram}: our sample.}
\label{fig1}
\end{figure}

The purpose of this work is to constrain the SFHs of these BCDs. However, as shown in Figure 2, the BCDs in our sample are all extended objects. The SDSS spectra are from the starlight that found its way down a $3''$-diameter fiber (with a physical size of $\sim 60$ pc at the nearest object Mrk 178 and $\sim 720$ pc at the most distant object SBS 1533+574) that have been positioned on bright H{\sc ii} regions for these BCDs (see Figure 2 for a direct view). Therefore, we need to check to what extent the SDSS fiber spectrum represents the light of the whole galaxy. 

\setcounter{figure}{1}
\begin{figure}[pthb]
\centering
\includegraphics[width=\textwidth]{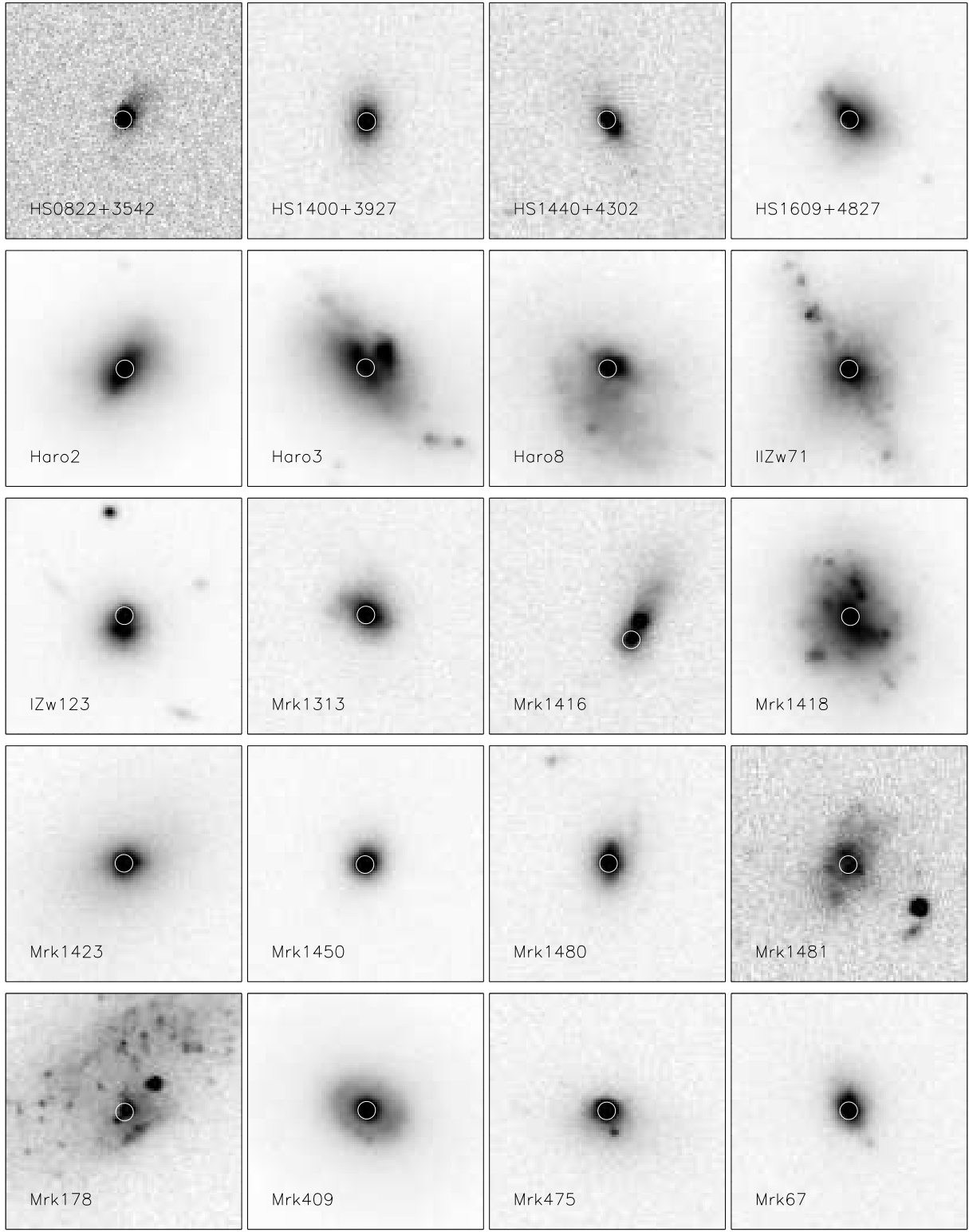}
\caption{SDSS $r$-band images of the BCDs. The overlaid circle in each panel shows the position of the SDSS $3''$-diameter fiber. The sizes of all images are $40''\times40''$.}
\label{Fig2}
\end{figure}

\setcounter{figure}{1}
\begin{figure}[t]
\centering
\includegraphics[width=\textwidth,bb=1 8 590 452]{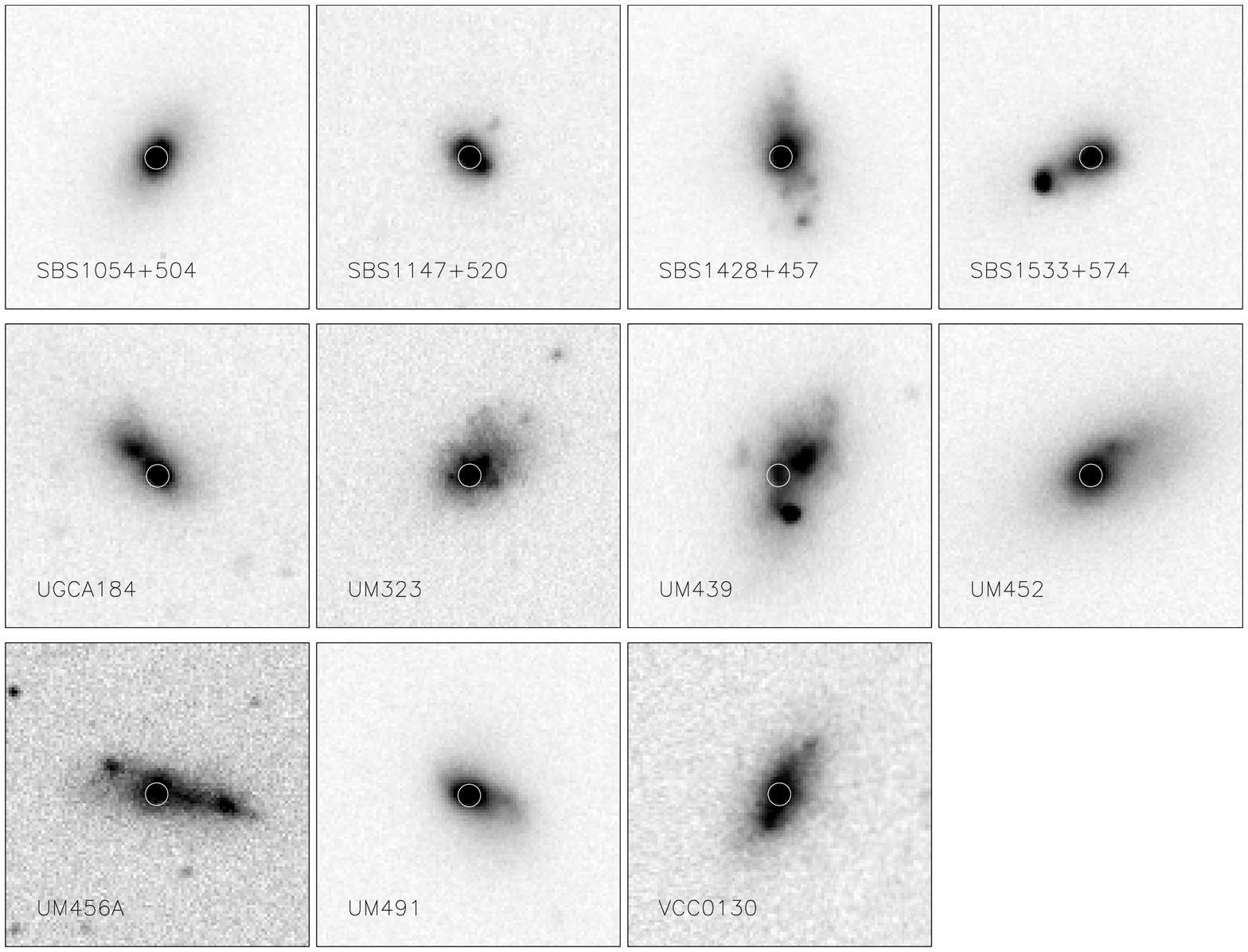}
\caption{Continued}
\label{Fig2cont}
\end{figure}

In Table 2, we list two quantities related to the covering fraction of the fixed-size aperture for all BCDs. One is the ratio ($r_{\rm{fiber}}/{\rm{petro}}R_{90}$) between the fiber radius ($1''.5$) and the Pertrosian radius containing 90 percent of the galaxy light in $r$-band, and the other is the light fraction, $f_r$, captured by the fiber. The $f_r$ is calculated using the total galaxy magnitude and the magnitude inside the fiber, explicitly $f_r=100\times10^{-0.4(m_{{\rm{fiber}}}-m_{{\rm{Pertro}}})}$. The median values of $r_{\rm{fiber}}/{\rm{petro}}R_{90}$ and $f_r$ are 0.37 and 14.4\%, respectively. For our sample galaxies, therefore, the SDSS spectrum could have much bluer $(g-r)$ color and less prominent 4000\AA\ break ($D_n(4000)$) (see below) than the integral spectrum. 

In Figure 3, we plot the observed SDSS spectra (black line) together with the integrated $ugriz$ broad-band spectral energy distributions (SEDs; red dashed line) for all galaxies in our sample. The flux has been normalized to the $i$-band value. For most galaxies, the continuum shape of the fiber spectrum longward of $r$-band is very consistent with that of the broad-band SED, whereas the spectrum shortward of $r$-band starts to deviate from the broad-band SED to varying degrees. It suggests that the aperture effects might much affect our fitting results presented in the next section for galaxies whose fiber spectra apparently differ from the broad-band SEDs  blueward of $r$-band. Fortunately, ten out of the 31 galaxies have been observed using drift-scanning technique (Moustakas \& Kennicutt 2006, hereafter MK06; private communication). The integrated spectra for these galaxies are superposed in Figure 3 with blue lines. Again, the aperture effects, especially of the 4000 \AA\ break, are clearly demonstrated. We will return to the question of how much the aperture effects will affect our results in the next and the discussion sections.

Kewley et al. (2005) have shown that the fiber needs to capture more than 20 percent of the galaxy light to minimize the aperture effects in the spectral measurements. According to this threshold we divide our galaxies into three subsamples: $f_r < 20\%$, $20\% \leqslant f_r <100\%$ and $f_r \simeq100\%$ (integrated spectra), and in the following we refer them to S1, S2 and S3, respectively. There are 16, 5 and 10 galaxies in S1, S2 and S3, respectively.

\begin{figure}[pthb]
\centering
\includegraphics[width=0.6\textwidth,bb=80 42 480 836]{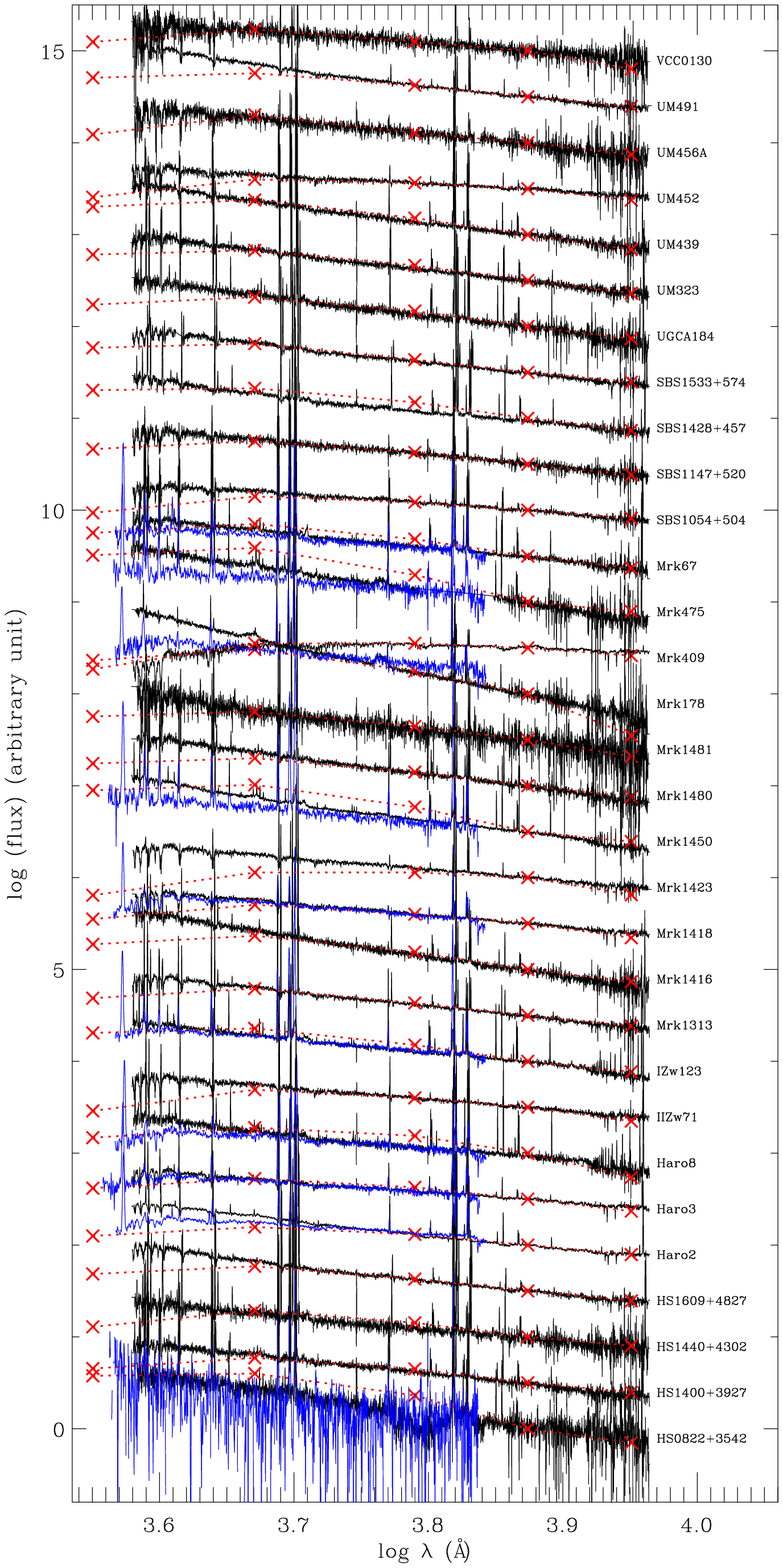}
\caption{The observed SDSS spectra (black line) and broad-band SEDs ($ugriz$; red dashed line) for these 31 BCDs. We also overplot the integrated spectra from MK06 (blue line) for 10 BCDs.}
\label{Fig3}
\end{figure}

\subsection{Stellar Population Synthesis}
In order to obtain the stellar populations for these BCDs and to correct for the stellar absorption in the emission lines, we here model the stellar contribution in the SDSS spectra through the modified version of the stellar population synthesis code, STARLIGHT\footnote{STARLIGHT \& SEAGal: http://www.starlight.ufsc.br/ } (Cid Fernandes et al. 2005 (C05), 2007;  Mateus et al. 2006; Asari et al. 2007), which adopted the stellar library from Bruzual \& Charlot (2003). The code does a search for the linear combination of SSPs to match a given observed spectrum ($O_\lambda$). The model spectrum ($M_\lambda$) is: 
\begin{center}
$M_\lambda =M_{\lambda_0}
   \left[
   \sum_{j=1}^{N_\star} x_j b_{j,\lambda} r_\lambda
   \right]
   \otimes G(v_\star,\sigma_\star)$,
 \end{center}
where $b_{j,\lambda} \equiv L_\lambda^{SSP}(t_j,Z_j) /L_{\lambda_0}^{SSP}(t_j,Z_j)$ is the spectrum of the $j^{\rm th}$ SSP normalized at $\lambda_0$, $r_\lambda \equiv 10^{-0.4(A_\lambda - A_{\lambda_0})}$ is the reddening term, {\boldmath $x$} is the population vector, $M_{\lambda_0}$ is the synthetic flux at the normalization wavelength, $N_\star$ is the total number of SSPs, $G(v_\star,\sigma_\star)$ is the line-of-sight stellar velocity distribution, modeled as a Gaussian centered at velocity $v_\star$ and broadened by $\sigma_\star$. The match between model and observed spectra is calculated by
\begin{center}
$\chi^2(x,M_{\lambda_0},A_V,v_\star,\sigma_\star) =
   \sum_{\lambda=1}^{N_\lambda}
   \left[
   \left(O_\lambda - M_\lambda \right) w_\lambda
   \right]^2$,
\end{center}
 where the weight spectrum $w_\lambda$ is defined as the inverse of the noise in $O_\lambda$. For more details, please refer to C05 and Mateus et al. (2006). The SSP library follows the work of SEAGal Group, and is made up of $N_\star=100$, including 25 ages (from 1 Myr to 18 Gyr) and 4 metallicities ($Z=0.005, 0.02, 0.2, {\rm and}\, 0.4 Z_\odot$, for our metal-poor sample). The spectra were computed with the Salpeter (1955) initial mass function (IMF), Padova 1994 models and the STELIB library (Le Borgne et al. 2003). The intrinsic reddening is modeled by the foreground dust model, using the extinction law of Calzetti et al. (1994) with $R_V = 4.05$ (Calzetti et al. 2000) for these star-forming galaxies.
 
The SDSS spectra cover 3800-9200 \AA, with a resolution ($\lambda/\Delta\lambda$) of $1800 < R < 2100$ and sampling of 2.4 pixels per resolution element. The fiber used in the SDSS spectroscopic observations has a diameter of 3$''$ on the sky. Prior to the synthesis, the Galactic extinction has been corrected with a combination of the extinction law of Cardelli et al. (1989) and the $A_B$ value from Schlegel et al. (1998) as listed in NED\footnote{http://nedwww.ipac.caltech.edu/}. The spectra are transformed into the rest frame using the redshifts given in the FITS header. The SSPs are normalized at $\lambda_0 = 4020$ \AA, while the observed spectra are normalized to the median flux between 4010 and 4060 \AA. The signal-to-noise ratio (S/N) is measured in the relatively clean window between 4730 and 4780 \AA. Masks of $20-30$ \AA\ around obvious emission lines are constructed for each object individually, and more weights are given to the strongest stellar absorption features such as Ca\,{\sc ii} K $\lambda$ 3934, and the Ca\,{\sc ii} triplets, that are less affected by nearby emission lines.

\begin{figure}[t]
\centering
\includegraphics[width=0.8\textwidth,bb=20 237 568 625]{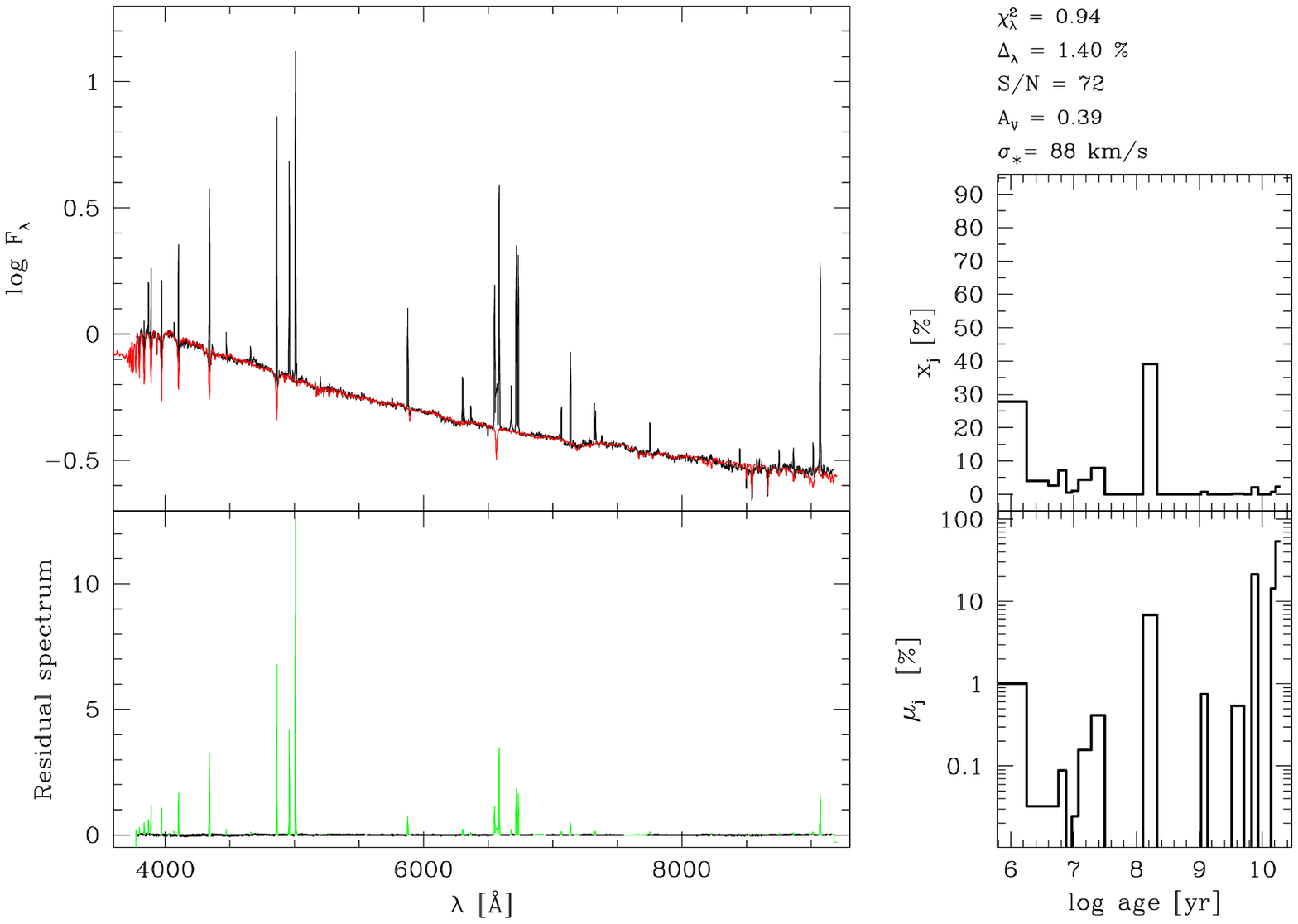}
\caption{Results of the spectral fitting for Haro 2. The top left panel shows the logarithm of the observed ($F_\lambda^O$; black line) and the synthetic ($F_\lambda^M$; red line) spectra. The $F_\lambda^O - F_\lambda^M$ residual spectrum is shown in the bottom left panel. Spectral regions actually used in the synthesis are plotted with a black line, while masked regions are plotted with a green line. Panels in the right show the population vector binned in the 25 ages of SSPs used in the model library. The top right panel corresponds to the population vector in flux fraction, normalized to $\lambda_0 = 4020$ \AA, while the corresponding mass fractions vector is shown in the bottom right panel.}
\label{Fig4}
\end{figure}

For our sample, the S/N varies between 6.8 and 72.0 (see Table 2), and most of them are larger than 20. Generally, the fitting results for high S/N objects are better than those for low S/N ones. Inspecting the fitting results, we find that the goodness of fitting ($\chi^2$ value) also somewhat depends on the absorption line equivalent widths (e.g. EW of Ca\,{\sc ii} K). A typical example of our fitting result for Haro 2 is shown in Figure 4. After subtracting the best-fit model spectrum from the observed one, we obtain the pure emission-line spectrum as shown in the bottom left panel of Figure 4, where we can derive accurate fluxes for all the emission lines with the {\it onedspec.splot} task in IRAF\footnote{IRAF is distributed by the National Optical Astronomical Observatory, which is operated by the Association of Universities for Research in Astronomy (AURA), Inc., under a cooperative agreement with the National Science Foundation.}.

\section{Results and analysis}
\subsection{Mean Stellar Age}

\begin{deluxetable}{lccccccc}
\tablewidth{0pt}
\tabletypesize{\scriptsize}
\tablecaption{Derived and fitting results for the BCD sample}
\tablehead{\colhead{}&\colhead{petro$R_{90}${\tablenotemark{b}}}&\colhead{}&\colhead{$f_r${\tablenotemark{c}}}&\colhead{}&\colhead{$\left <\log t_\star \right >_L$}&\colhead{$\left <\log t_\star \right >_M$}\\
\colhead{Galaxy}&\colhead{(arcsec)}&\colhead{$R_{\rm{fiber}}/R_{90}$}&\colhead{(\%)}&\colhead{S/N{\tablenotemark{d}}}&\colhead{(yr)}&\colhead{(yr)}}
\startdata
HS 0822+3542{\tablenotemark{a}}\dotfill & 1.9&0.80&49.5& 6.8& 7.68& 8.09\\
&&&&2.6&7.10&9.71\\
HS 1400+3927\dotfill &  3.0&0.51&25.3&22.7& 7.40& 8.27\\
HS 1440+4302\dotfill &  2.9&0.52&26.7&12.2& 7.43& 9.85\\
HS 1609+4827\dotfill &  3.9&0.39&13.6&34.4& 7.23& 9.87\\
Haro 2{\tablenotemark{a}}\dotfill &  5.8&0.26&13.4&72.1& 7.36& 9.94\\
&&&&62.2&7.76&9.91\\
Haro 3{\tablenotemark{a}}\dotfill &  7.4&0.20& 5.0&52.2& 7.76& 9.56\\
&&&&31.0&7.99 &9.94\\
Haro 8{\tablenotemark{a}}\dotfill &  8.9&0.17& 7.2&25.9& 7.47&10.02\\
&&&&21.8&8.32&9.89\\
I Zw 123{\tablenotemark{a}}\dotfill &  3.7&0.41&17.2&12.1& 6.99& 9.98\\
&&&&34.8&7.19&9.53\\
II Zw 71\dotfill &  8.8&0.17& 4.3&46.0& 7.89& 9.34\\
Mrk 1313\dotfill &  3.1&0.48&19.3&36.4& 7.68& 8.92\\
Mrk 1416\dotfill &  4.5&0.34&15.3&25.7& 6.87& 7.10\\
Mrk 1418{\tablenotemark{a}}\dotfill &  7.0&0.21& 3.7&49.0& 7.96& 8.86\\
&&&&49.6&7.94& 9.90\\
Mrk 1423\dotfill & 11.2&0.13& 6.4&33.5& 7.75& 8.67\\
Mrk 1450{\tablenotemark{a}}\dotfill &  2.1&0.70&33.6&20.5& 7.08& 9.96\\
&&&&17.7&7.03& 9.74\\
Mrk 1480\dotfill &  3.2&0.47&22.6&31.4& 7.01& 9.58\\
Mrk 1481\dotfill &  6.6&0.23& 5.8& 7.7& 7.37& 9.87\\
Mrk 178{\tablenotemark{a}}\dotfill &  8.4&0.18& 7.3&34.0& 6.29& 6.37\\
&&&&25.8&7.43 &9.67\\
Mrk 409\dotfill &  6.5&0.23&14.4&28.0& 9.00& 9.87\\
Mrk 475{\tablenotemark{a}}\dotfill &  2.5&0.59&28.8&14.7& 6.68& 6.64\\
&&&&15.5&7.34& 9.81\\
Mrk 67{\tablenotemark{a}}\dotfill &  3.3&0.46&22.4&18.2& 7.08& 9.52\\
&&&&25.0&7.38&9.87\\
SBS 1054+504\dotfill &  4.0&0.37&20.1&32.9& 7.97& 9.51\\
SBS 1147+520\dotfill &  2.5&0.61&24.0&21.0& 7.83& 8.94\\
SBS 1428+457\dotfill &  4.3&0.35&15.3&35.4& 6.99&10.04\\
SBS 1533+574\dotfill &  3.7&0.41&18.5&36.2& 7.36& 8.05\\
UGCA 184\dotfill &  4.9&0.31&11.8&13.6& 6.67& 9.75\\
UM 323\dotfill &  3.4&0.44&12.8&26.0& 7.28& 7.70\\
UM 439\dotfill &  6.5&0.23&12.6&27.5& 7.10& 9.75\\
UM 452\dotfill &  6.9&0.22& 7.6&30.0& 7.75& 9.78\\
UM 456A\dotfill &  4.1&0.36&10.5&13.4& 8.19& 9.48\\
UM 491\dotfill &  3.9&0.38&17.6&44.2& 7.17& 9.91\\
VCC 0130\dotfill &  3.9&0.39& 9.0& 8.6& 8.13& 9.52\\
\enddata
\label{table2}
\tablenotetext{a}{The second line gives the fitting results based on the integrated spectra.}
\tablenotetext{b}{The Petrosian radius containing 90 percent of the galaxy flux in $r$-band.}
\tablenotetext{c}{The fraction of the $r$-band flux covered by the SDSS fiber.}
\tablenotetext{d}{The signal-to-noise in the S/N window.}
\end{deluxetable}

In the top panels of Figure 5, we show the distributions of the mean stellar ages (C05) estimated for all of the three subsamples (Figure 5$a$, 5$b$ and 5$c$ for S1, S2 and S3, respectively). They are the mass-weighted mean stellar age,  
\begin{equation}
\left<\log\, t_\star \right>_M = \sum_{j=1}^{N_\star} \mu_j\, \log\, t_j,
\end{equation}
and the light-weighted mean stellar age,
\begin{equation}
\left<\log\, t_\star \right>_L = \sum_{j=1}^{N_\star} x_j\, \log\, t_j.
\end{equation}
where $\mu_j$ and $x_j$ represent the fractional contributions to the stellar mass and luminosity of the SSP with age $t_j$ respectively. $N_\star$ is the number of SSPs. Based on these two definitions, it is easy to understand that $\left< t_\star \right>_M$ represents the epoch of the stellar population formation and these stars nowadays contribute most significantly to the stellar mass in galaxies. Consequently, $\left<t_\star\right>_M$ is associated with the mass assembly history of galaxies. On the other hand, $\left<t_\star \right>_L$ reflects the epoch of the formation of massive and bright O and B stars, frequently associated with recent starburst activities. Therefore, $\left<t_\star \right>_L$ is strongly affected by the recent SFH of a given galaxy. The fitting results of these two kinds of ages for our BCD sample are summarized in Table 2. According to C05 and  Mateus et al. (2006), the uncertainties of these two parameters depend on the S/N of the input spectra. In general, the rms of the fitted $\left<\log\, t_\star \right>_M$ is $\sim 0.2$ dex for $\rm{S/N} < 10$ and $\sim 0.1$ dex for $\rm{S/N} > 10$, while the rms of the fitted $\left<\log\, t_\star \right>_L$ is $\sim 0.15$ dex for $\rm{S/N} \sim 5$ and $< 0.1$ dex for $\rm{S/N} > 10$.

In each panel of the top row of Figure 5, the solid line shows $\left<\log t_\star \right>_M$, while the dashed line shows $\left<\log t_\star \right>_L$. For all of the three subsamples, $\left<\log t_\star \right>_L$ is in the range of several Myr to a few 100 Myr, with a trend of the median $\left<\log t_\star \right>_L$ towards larger value from S1 to S3. 

\begin{figure}[phtb]
\centering
{\includegraphics[bb=20 38 500 330,width=0.8\textwidth]{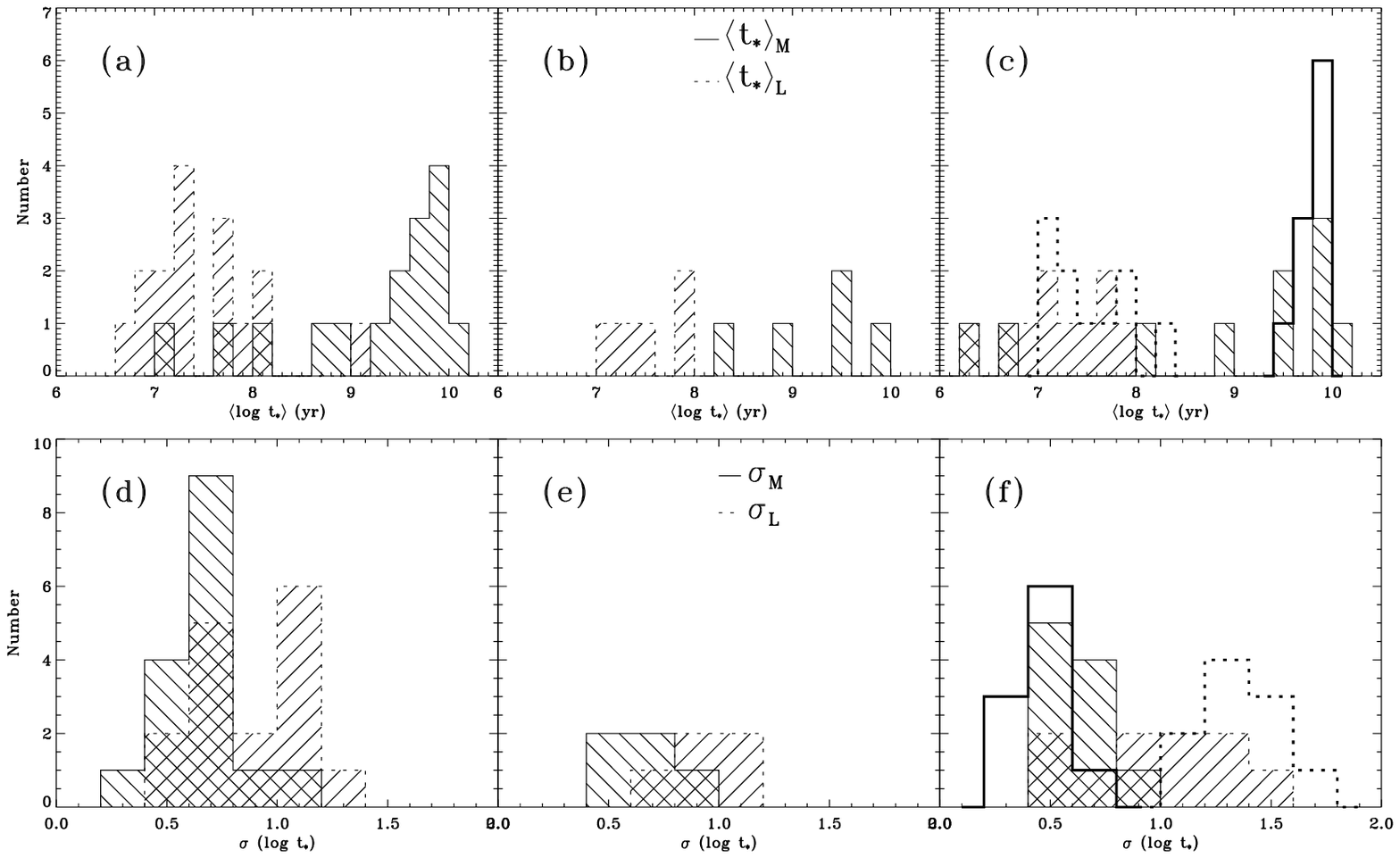}}
\caption{Statistics on the fitting results (from left to right for subsamples: S1, S2, and S3). In each panel, the mass-weighted and light-weighted results are shown with solid and dashed lines, respectively. The filled and (thick) blank histograms in the right panels ($c$ and $f$) show the results derived with the fiber and integrated spectra, respectively. {\it Top panels}: Number distributions of the BCDs in mean stellar ages. {\it Bottom panels}: Number distributions of the BCDs in the dispersions of $\left<\log t_\star\right>$.}
\label{Fig5}
\end{figure}

As shown in Figure 5$a$, for the S1 sample, whose spectra might severely under-sample the member galaxies, $\left<\log t_\star \right>_M$ ranges from $\sim 10$ Myr to $\sim 10$ Gyr, and 11 out 16 galaxies have $\left<\log t_\star \right>_M > 1$ Gyr. For the S2 sample, whose spectra may be representative of the member galaxies, $\left<\log t_\star \right>_M$ is in the range of several 100 Myr to 10 Gyr (Figure 5$b$). For the S3 sample, which have integrated spectra, all of $\left<\log t_\star \right>_M$ are much larger than 1 Gyr (blank, thick histogram in Figure 5$c$). These results indicate that the aperture bias will lead to an underestimate of $\left<\log t_\star \right>_M$, which can also be seen from Figure 5$c$, in which the filled histogram shows the results derived from the fiber spectra. To check the aperture effect further, we compare results derived from the integrated spectra with those from the fiber spectra for the 4 galaxies having $\left<\log t_\star \right>_M < 1$ Gyr, and find that the mean values of $\Delta \left<\log t_\star \right>_L(\equiv \left<\log t_\star \right>_{L,\,{\rm integrated}}-\left<\log t_\star \right>_{L,\,3''})$  and $\Delta \left<\log t_\star \right>_M(\equiv \left<\log t_\star \right>_{M,\,{\rm integrated}}-\left<\log t_\star \right>_{M,\,3''})$ are $0.34\pm0.49$ and $0.90\pm1.38$, respectively. Thus, while both $\left<\log t_\star \right>_L$ and $\left<\log t_\star \right>_M$ are underestimated based on the fiber spectra, $\left<\log t_\star \right>_M$ suffers much more severely from the aperture effects.

In our sample, Mrk 178 is the only galaxy which has been studied extensively using CMDs in literature. Schulte-Ladbeck et al. (2000) found RGB stars with age of at least 1 Gyr in this galaxy and concluded that this galaxy should be an old galaxy. The $\left<\log t_\star \right>_M$ derived from the integrated spectrum is as old as $\sim 5$ Gyr, and is well consistent with the results of Schulte-Ladbeck et al. (2000). 

In summary, although a part of our sample may suffer from the aperture effects, our results are generally consistent with the findings through deep photometric observations for nearby BCDs (e.g. Schulte-Ladbeck et al. 2001a; Aloisi et al. 2007), and clearly show that BCDs are not truly primordial galaxies, rather, they are old systems undergoing transient periods of starburst. 

To investigate the SFH of BCDs in more detail, we calculated the light-weighted and mass-weighted standard deviations of the log age, which are defined as the following (C05),
\begin{equation}
\sigma _L (\log t_\star ) = \left[ {\sum\limits_{j = 1}^{N_\star } {x_j (\log t_j  - \left\langle {\log t_\star } \right\rangle _L )^2 } } \right]^{1/2}. 
\end{equation}
and
\begin{equation}
\sigma _M (\log t_\star ) = \left[ {\sum\limits_{j = 1}^{N_\star } {\mu_j (\log t_j  - \left\langle {\log t_\star } \right\rangle _M )^2 } } \right]^{1/2}.
\end{equation}
These two higher moments of the age distribution could be used to distinguish galaxies dominated by a single population from those which
had bursty or continuous SFHs.  

The bottom panels in Figure 5 displays the number frequency histograms of BCDs in these two parameters. Both of the median values of $\sigma _M (\log t_\star )$ and $\sigma _L (\log t_\star )$ are significantly larger than zero, indicating that BCDs are not dominated by a single population. $\sigma _L (\log t_\star )$ has a larger median value than $\sigma _M (\log t_\star )$. This suggests that the majority of the stellar mass in BCDs is assembled in a relatively narrow time range, while the star formation activities might occur repeatedly during the lifetime of a BCD. At the same time, we can see from Figure 5$f$ that the aperture effects generally result in an underestimate of $\sigma _L (\log t_\star )$ and an overestimate of $\sigma _M (\log t_\star )$.

\subsection{Luminosity and mass fractions of different populations}

\begin{figure}[pthb]
\centering
\includegraphics[width=0.8\textwidth, bb=60 1 450 360]{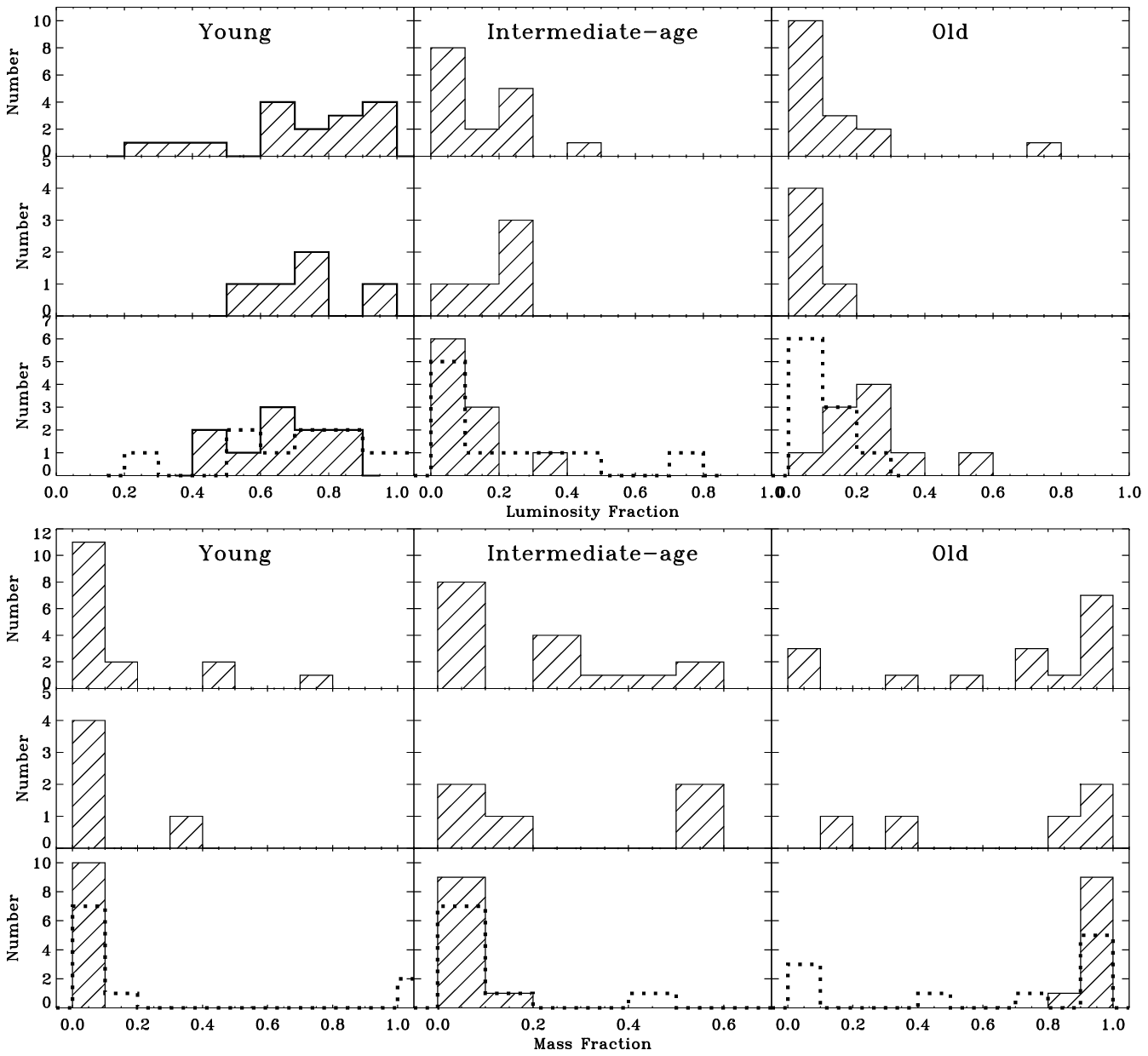}
\caption{Number distributions of the BCDs in luminosity (top half page) and mass (bottom half page) fractions for the young ($t<10^8$ yr), intermediate ($10^8\leq t\leq 10^9$ yr) and old ($t >10^9$ yr) stellar populations. Solid and dashed histograms show the integrated and fiber spectra-based results, respectively. For each of the half page, from top to bottom: S1, S2, and S3.}
\label{Fig6}
\end{figure}

A coarse but robust description of the SFH of a galaxy may be obtained by binning {\boldmath $x$ ($\mu$)} into `young' ($t_j < 10^8$ yr), `intermediate-age'  ($10^8 \leq t_j \leq 10^9$ yr), and `old' ($t_j > 10^9$ yr) components  ($x_{\rm Y}$ ($\mu_{\rm Y}$), $x_{\rm I}$ ($\mu_{\rm I}$), and $x_{\rm O}$ ($\mu_{\rm O}$), respectively). As shown in Figure 6, our stellar synthesis result indicates that the luminosity and stellar mass of the BCDs are contributed by different components. For most galaxies, the young population contributes the largest fraction of the 4020 \AA\ monochromatic radiation, while the old population contributes most stellar mass. 

For the S3 sample, more than 50\% of the 4020 \AA\ monochromatic radiation comes from the young population for 8 out of 10 galaxies, while more than 80\% of the stellar mass is contributed by the old population for all of the galaxies. For the S1 and S2 samples, which might suffer severely from aperture effects, the young and inter-mediate populations have more contributions to the luminosity and mass fractions than the S3 sample. About $1/3-1/2$ galaxies in S1 and S3 have more than 20\% of their stellar mass and light contributed by the intermediate-age population.  Whereas for the S3 sample, the intermediate-age population only contributes $<20\%$ light and stellar mass for all of the galaxies (only one exception that the intermediate-age population contributes 30-40\% light).

Our results (specially from the integrated spectra) are in agreement with the scenario that BCDs are old galaxies with episodic starbursting activities. The scarcity of intermediate-age component suggests that BCDs should not have a SFH with a constant (current) SFR. However, we can not discriminate the scenario that there exists extensively long quiescent phases during a BCD galaxy lifetime from that a BCD is still forming stars with a very low SFR during its ``quiescent" periods.

\subsection{Extinction and the Age-Extinction Degeneracy}
There are several independent ways to measure the internal extinctions. In this paper, we use the continuum and nebular lines to make separate estimates of the internal extinctions in the BCDs. {\sc starlight} also returns an estimate of the stellar visual extinction, $A^\star_V$, modeled as due to a foreground dust screen.

By making the reasonable assumption that the intrinsic Balmer line ratios are equal to the Case B recombination and using a Calzetti et al. (1994) reddening law, we could estimate the nebular extinctions from the observed ${\rm{H}}\alpha/{\rm{H}}\beta$ \ and ${\rm{H}}\gamma/{\rm{H}}\beta$\ Balmer decrements, which are

 \begin{equation}
 A_{V}^{{\rm{H}}\alpha} = 7.93 \times \log
(\frac{F_{{\rm{H}}\alpha}/F_{{\rm{H}}\beta}}{I_{{\rm{H}}\alpha}/I_{{\rm{H}}\beta}})
 \end{equation}

\begin{equation}
 A_{V}^{{\rm{H}}\gamma} = -19.04 \times \log
(\frac{F_{{\rm{H}}\gamma}/F_{{\rm{H}}\beta}}{I_{{\rm{H}}\gamma}/I_{{\rm{H}}\beta}})
 \end{equation}
where $F_{{\rm{H}}\alpha}/F_{{\rm{H}}\beta}$,$F_{{\rm{H}}\gamma}/F_{{\rm{H}}\beta}$ and $I_{{\rm{H}}\alpha}/I_{{\rm{H}}\beta}$, $I_{{\rm{H}}\gamma}/I_{{\rm{H}}\beta}$ are the observed and intrinsic Balmer decrements, respectively. Here we adopt the intrinsic ratios of $I_{{\rm{H}}\alpha}/I_{{\rm{H}}\beta}$ and $I_{{\rm{H}}\gamma}/I_{{\rm{H}}\beta}$ to be 2.86 and 0.469 (Brocklehurst 1971), respectively, for an electron temperature of 10$^4$ K and an electron density of 100 cm$^{-3}$. We prefer using equation (5) to calculate the extinction since ${\rm{H}}\gamma$ has relatively lower S/N than ${\rm{H}}\alpha$ and equation (6) is much more sensitive to the measurement errors compared to equation (5). The observed fluxes of H$\alpha$ and H$\beta$ are measured through the ``pure-emission'' spectra, and are presented in Table 3. The errors in the line fluxes are calculated according to files attached to each spectrum which contain the error in each pixel. Additionally, an error of $\sim 2\%$ in the emission line fluxes should be accounted for the uncertainties of standard star absolute fluxes used for flux calibration (e.g., Oke 1990). For sources with saturated H$\alpha$, the adopted extinction is calculated using equation (6). The $A_V$ values calculated through Balmer decrements are given in Table 3. 

\begin{deluxetable}{lcccccccc}
\tabletypesize{\scriptsize}
\tablewidth{0pt}
\tablecaption{Observed and derived properties for the BCD sample}
\tablehead{
&\multicolumn{2}{c}{Observed flux\tablenotemark{a}}&\colhead{$A_V$\tablenotemark{b}}&\colhead{}&\colhead{SFR\tablenotemark{d}}&\colhead{$R_{\rm{H}\alpha}$\tablenotemark{e}}&\colhead{SFR$_{tot}$\tablenotemark{f}}\\
\cline{2-3}
\colhead{Galaxy}&\colhead{H$\beta$}&\colhead{H$\alpha$}&\colhead{(mag)}&\colhead{$L_{\rm{H}\alpha}$\tablenotemark{c}}&\colhead{($M_\odot$ yr$^{-1}$)}&\colhead{(\%)}&\colhead{($M_\odot$ yr$^{-1}$)}}
\startdata
HS 0822+3542\dotfill&$2.425\pm0.020$E+3&$7.400\pm0.055$E+3&0.22&1.07E$+$1&8.45E$-$3&101.4&8.33E$-$3\\
HS 1400+3927\dotfill&$1.661\pm0.014$E+3&$4.747\pm0.032$E+3&0.00&2.53E$+$1&2.00E$-$2& 55.4&3.61E$-$2\\
HS 1440+4302\dotfill&$1.116\pm0.011$E+3&$3.452\pm0.027$E+3&0.27&7.32E$+$1&5.79E$-$2& 84.3&6.86E$-$2\\
HS 1609+4827\dotfill&$1.277\pm0.015$E+3&$4.072\pm0.029$E+3&0.37&1.17E$+$2&9.28E$-$2& 19.4&4.80E$-$1\\
                 Haro 2\dotfill&$1.677\pm0.013$E+4&\nodata&0.48&4.45E$+$2&3.51E$-$1& 35.4&9.92E$-$1\\
      Haro 3\dotfill&$4.502\pm0.040$E+3&$1.348\pm0.009$E+4&0.16&3.77E$+$1&2.98E$-$2&  4.9&6.02E$-$1\\
                Haro 8\dotfill&$5.552\pm0.047$E+3&\nodata& 0.62&9.31E$+$1&7.36E$-$2& 44.2&1.66E$-$1\\
              I Zw 123\dotfill&$7.730\pm0.070$E+3&\nodata& 2.20&3.62E$+$2&2.86E$-$1&119.0&2.40E$-$1\\
   II Zw 71\dotfill&$4.916\pm0.093$E+2&$1.647\pm0.013$E+3& 0.54&1.04E$+$1&8.22E$-$3&  4.1&2.03E$-$1\\
   Mrk 1313\dotfill&$1.473\pm0.011$E+3&$4.406\pm0.027$E+3& 0.15&6.00E$+$1&4.74E$-$2& 50.9&9.30E$-$2\\
   Mrk 1416\dotfill&$1.085\pm0.011$E+3&$3.319\pm0.019$E+3& 0.23&5.41E$+$1&4.27E$-$2& 17.3&2.48E$-$1\\
   Mrk 1418\dotfill&$4.173\pm0.105$E+2&$1.255\pm0.012$E+3& 0.17&2.23E$+$0&1.76E$-$3&  1.3&1.33E$-$1\\
   Mrk 1423\dotfill&$5.259\pm0.073$E+2&$1.934\pm0.010$E+3& 0.87&1.86E$+$1&1.47E$-$2& 28.4&5.15E$-$2\\
              Mrk 1450\dotfill&$1.201\pm0.010$E+4&\nodata& 0.22&1.12E$+$2&8.81E$-$2& 86.3&1.02E$-$1\\
   Mrk 1480\dotfill&$2.404\pm0.021$E+3&$7.680\pm0.059$E+3& 0.38&9.08E$+$1&7.17E$-$2& 55.3&1.30E$-$1\\
   Mrk 1481\dotfill&$2.595\pm0.055$E+2&$7.640\pm0.089$E+2& 0.10&7.35E$+$0&5.80E$-$3& 17.1&3.39E$-$2\\
    Mrk 178\dotfill&$5.877\pm0.082$E+2&$1.830\pm0.013$E+3& 0.29&4.82E$-$1&3.81E$-$4&  1.7&2.26E$-$2\\
    Mrk 409\dotfill&$6.491\pm0.189$E+2&$2.152\pm0.026$E+3& 0.51&1.72E$+$1&1.36E$-$2&  7.4&1.84E$-$1\\
               Mrk 475\dotfill&$4.238\pm0.033$E+3&\nodata& 0.16&1.61E$+$1&1.27E$-$2& 35.9&3.54E$-$2\\
                Mrk 67\dotfill&$3.831\pm0.043$E+3&\nodata& 0.26&3.51E$+$1&2.77E$-$2& 46.8&5.91E$-$2\\
SBS 1054+504\dotfill&$1.057\pm0.013$E+3&$3.542\pm0.024$E+3&0.55&2.59E$+$1&2.04E$-$2& 82.0&2.49E$-$2\\
SBS 1147+520\dotfill&$3.539\pm0.071$E+2&$1.233\pm0.010$E+3&0.68&8.82E$+$0&6.97E$-$3& 87.5&7.96E$-$3\\
SBS 1428+457\dotfill&$5.190\pm0.045$E+3&$1.699\pm0.010$E+4&0.47&3.61E$+$2&2.85E$-$1& 35.8&7.95E$-$1\\
SBS 1533+574\dotfill&$1.064\pm0.013$E+3&$3.248\pm0.023$E+3&0.22&1.13E$+$2&8.96E$-$2& 10.7&8.34E$-$1\\
    UGCA 184\dotfill&$2.841\pm0.021$E+3&$8.861\pm0.061$E+3&0.30&7.03E$+$1&5.55E$-$2& 63.0&8.82E$-$2\\
                UM 323\dotfill&$1.309\pm0.013$E+3&\nodata& 0.01&3.61E$+$1&2.86E$-$2& 26.7&1.07E$-$1\\
     UM 439\dotfill&$9.399\pm0.089$E+2&$2.976\pm0.017$E+3& 0.35&9.10E$+$0&7.19E$-$3&  7.0&1.03E$-$1\\
     UM 452\dotfill&$5.369\pm0.082$E+2&$1.942\pm0.015$E+3& 0.81&1.63E$+$1&1.29E$-$2& 41.4&3.11E$-$2\\
    UM 456A\dotfill&$4.501\pm0.297$E+1&$1.599\pm0.023$E+2& 0.75&2.02E$+$0&1.60E$-$3&  3.8&4.20E$-$2\\
     UM 491\dotfill&$9.774\pm0.122$E+2&$3.029\pm0.025$E+3& 0.28&3.41E$+$1&2.69E$-$2& 34.8&7.74E$-$2\\
   VCC0 130\dotfill&$5.543\pm0.433$E+1&$1.873\pm0.034$E+2& 0.57&8.86E$-$1&7.00E$-$4& 10.7&6.55E$-$3\\
\enddata
\tablenotetext{a}{In units of $10^{-17} \mathrm{erg~s}^{-1}~\mathrm{cm}^{-2}$.}
\tablenotetext{b}{Internal extinction derived with the flux ratio of Balmer lines.}
\tablenotetext{c}{Internal extinction-corrected luminosity of H$\alpha$, in units of 10$^{38}$ erg s$^{-1}$. For spectra that the H$\alpha$ lines are saturated, we use 2.86$\times L_{\rm{H\beta}}$ to estimate the H$\alpha$ luminosity.}
\tablenotetext{d}{Star formation rate, derived with the H$\alpha$ luminosity in Col (5)}
\tablenotetext{e}{The ratio (expressed as a percentage) of our H$\alpha$ luminosity to G03's}
\tablenotetext{f}{Total star formation rate, calculated based on the internal extinction-corrected \Ha\ luminosity given in G03.}
\label{table3}
\end{deluxetable}

We compare these two extinctions for our BCD sample, and find that $A^\star_V$, which is returned by {\sc starlight}, in general, is comparable with $A_{V,\,{\rm{neb}}}$. The median value of $A_{V,\,{\rm{neb}}}/A^\star_V$ is about 1.0, which is different from that found in the detailed studies of nearby star forming galaxies by Calzetti et al. (1994), who postulate that nebular line emission is attenuated by roughly twice as much dust as the stellar continuum (see also Charlot \& Fall 2000). This result may be due to the well known age-extinction degeneracy (e.g. Gordon et al. 1997), which acts in the sense of confusing old, less dusty systems with young, dusty ones and vice versa.

 The age-extinction degeneracy will affect the following fitted parameters, e.g. the luminosity- and mass-weighted mean stellar ages, the luminosity and mass fractions of different populations, and the stellar mass, to varying degrees, and we will further discuss the uncertainties caused by this effect in Section 4.1.

\subsection{Stellar Mass}
Through the spectral synthesis, {\sc starlight} presents the current stellar mass, $M_\star$, which correspond to the light entering the fiber. The total stellar masses of the galaxies are obtained from $M_\star$ divided by $f$, where $f$ is the fraction of the total galaxy luminosity in the $i$-band inside the fiber. For most galaxies, this can give a good estimate of the total stellar mass. However, this method will significantly underestimate the stellar mass for galaxies whose fiber spectra badly represent the whole galaxy in our sample. Therefore, for galaxies having integrated spectra we prefer to using the stellar masses derived from the integrated spectra. For galaxies whose $A^\star_{V,\,{\rm neb}}$ are significantly overestimated and which do not have integrated spectra, the stellar masses are derived by re-running {\sc starlight} with the same parameters as used in Section 2.2 except that we limit $A^\star_V$ to the value of $A_{V,\,{\rm{neb}}}$ (see Section 4.1). After these procedures, the stellar mass of three galaxies (Mrk 1416, Mrk 1423 and SBS1147+520) might be still much underestimated since their fitted $\left<\log t_\star\right>_M$ are all a bit less than 1 Gyr. We remove these three galaxies from further analysis involving stellar mass.

\subsection{Star formation properties}
Using the relation between SFR and nebular emission line (H$\alpha$) luminosity (Kennicutt 1998), namely
\begin{center}
\begin{equation}
{\rm{SFR}}(M_\odot\, {\rm{year}}^{-1})=7.9\times10^{-42} L({\rm{H}}\alpha)\,({\rm{ergs\,s}}^{-1}),
\end{equation}
\end{center}
we can estimate SFR for these BCDs.  Prior to the conversion, the internal extinction is corrected by using the attenuations calculated in the previous section using the Balmer line ratios and assuming the Calzetti et al. (1994) reddening law.  The H$\alpha$ luminosity and SFR covered by the SDSS fiber are listed in Table 3, Columns 5 and 6, respectively.

We also used the H$\alpha$ luminosities measured by G03, after the correction of the internal reddening using the $A_{V,\,{\rm neb}}$ values measured here, to calculate the integrated SFRs of the BCD sample. This is because that the G03 measurements of the H$\alpha$ luminosity are based upon narrowband imaging (having been corrected for the [N\,{\sc ii}] contribution; please see the Appendix in G03) and do not suffer from the aperture effects that plague the spectral fluxes. The integrated SFRs are given in Table 3, Column 8.  In the following, parameters or statistics relating to SFR are all based on the {\em integrated} SFRs unless otherwise stated. 

\begin{figure}[t]
\centering
\includegraphics[width=0.8\textwidth]{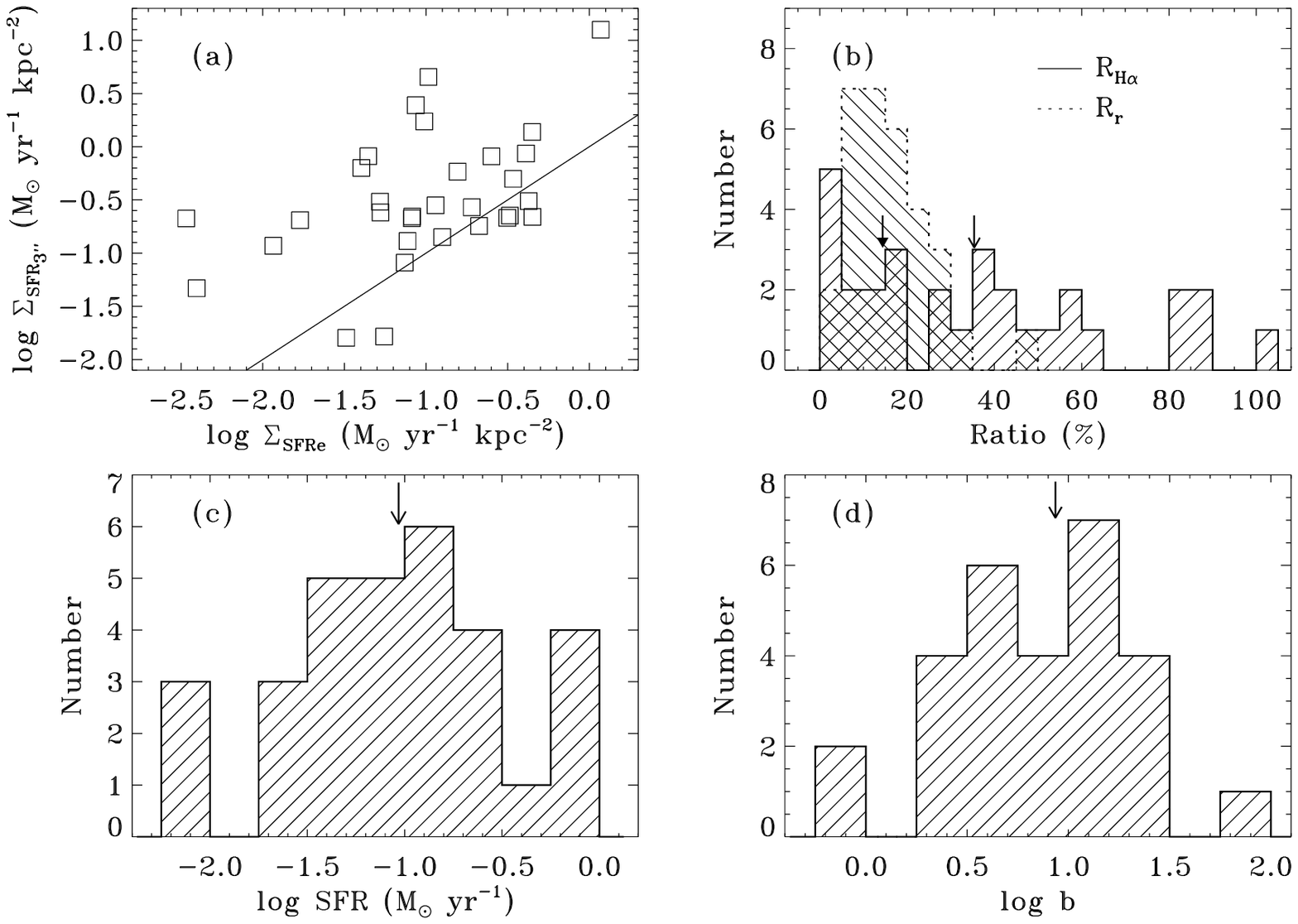}
\caption{ Panel (a) shows the correlation of SFR surface density between $\Sigma_{{\rm SFR}_{3''}}$ and $\Sigma_{{\rm SFR}_{\rm e}}$. The other three panels plot the number distributions of the BCDs in (b) fractions of the H$\alpha$ (solid) and $r$-band (dashed) fluxes inside the fiber, (c) star formation rate, and (d) birthrate parameter, $b$. The solid line in panel (a) is a reference line for the case when the two quantities are the same. The arrow in each panel marks the position of the median value.}
\label{Fig7}
\end{figure}

In Figure 7 we demonstrate the integrated and $3''$-sized star formation properties of these 31 BCDs. Figure 7$a$ illustrates the correlation between the SFR surface densities. The $x$-axis, $\Sigma_{{\rm SFR}_{\rm e}}$, is the integrated SFR normalized to the area defined by the scale length measured from exponential fits to $R$-band images (Gil de Paz \& Madore 2005). The $y$-axis, $\Sigma_{{\rm SFR}_{3''}}$, is the SFR surface density measured within the SDSS $3''$-diameter aperture. On one hand, these two parameters show an expected correlation that a galaxy with higher $\Sigma_{{\rm SFR}_{\rm e}}$ also has higher $\Sigma_{{\rm SFR}_{3''}}$. On the other hand, $\Sigma_{{\rm SFR}_{3''}}$ is larger than $\Sigma_{{\rm SFR}_{\rm e}}$ for most galaxies in this BCD sample, which is due to that these BCDs are extended objects while the SDSS fiber targets at the brightest part in most cases. As shown by the solid histogram in Figure 7$b$, the fraction of the H$\alpha$ flux ($R_{{\rm H}\alpha}$; see Table 3) captured in the fiber spectrum is small for most galaxies, with a median value of 35.4\%. For nearly one fourth of the galaxies, $R_{{\rm H}\alpha}$ is less than $\sim 10\%$. However, the median value of $R_{{\rm H}\alpha}$ is much larger than that of $R_{r}(\equiv f_r)$ (see the arrows in Figure 7$b$), which indicates that the distribution of H$\alpha$ flux is relatively more concentrated.

Figure 7$c$ shows the number distribution of the SFRs for these 31 BCDs. More than half galaxies in this sample have SFRs less than 0.1 $M_\odot$ yr$^{-1}$. The median value of the SFRs is 0.093 $M_\odot$ yr$^{-1}$.  Using the stellar mass derived in Section 3.4, we can determine the birthrate parameter (Kennicutt 1983), $b$, for this sample of BCDs. The parameter $b$ is defined as the ratio of the current SFR to the average past value, ${\rm{SFR/<SFR>_{past}}}$, where $<$SFR$>_{\rm{past}}$ is defined as the stellar mass, $M_{\star}$, divided by the stellar age, $t$. To minimize the aperture effect, the stellar age we adopted here is the age of the oldest population given by the fitting procedure.

Figure 7$d$ displays the distribution of $\log b$. The $b$ parameter is usually used to separate starburst galaxies from normal star forming galaxies. A galaxy with $b>2\sim3$ is defined as a starburst galaxy (e.g., Hunter \& Gallagher 1986; Salzer 1989; Gallagher 2005; Brinchmann et al. 2004; Kennicutt et al. 2005). For the subsample, the median value of $\log b$ is about  0.9 and more than 90 percent of the BCDs (26/28) have $\log b \geq  0.3$, which indicates an ongoing starburst activity.

\begin{figure}[t]
\centering
{\includegraphics[bb=104 4 377 356,width=0.5\textwidth]{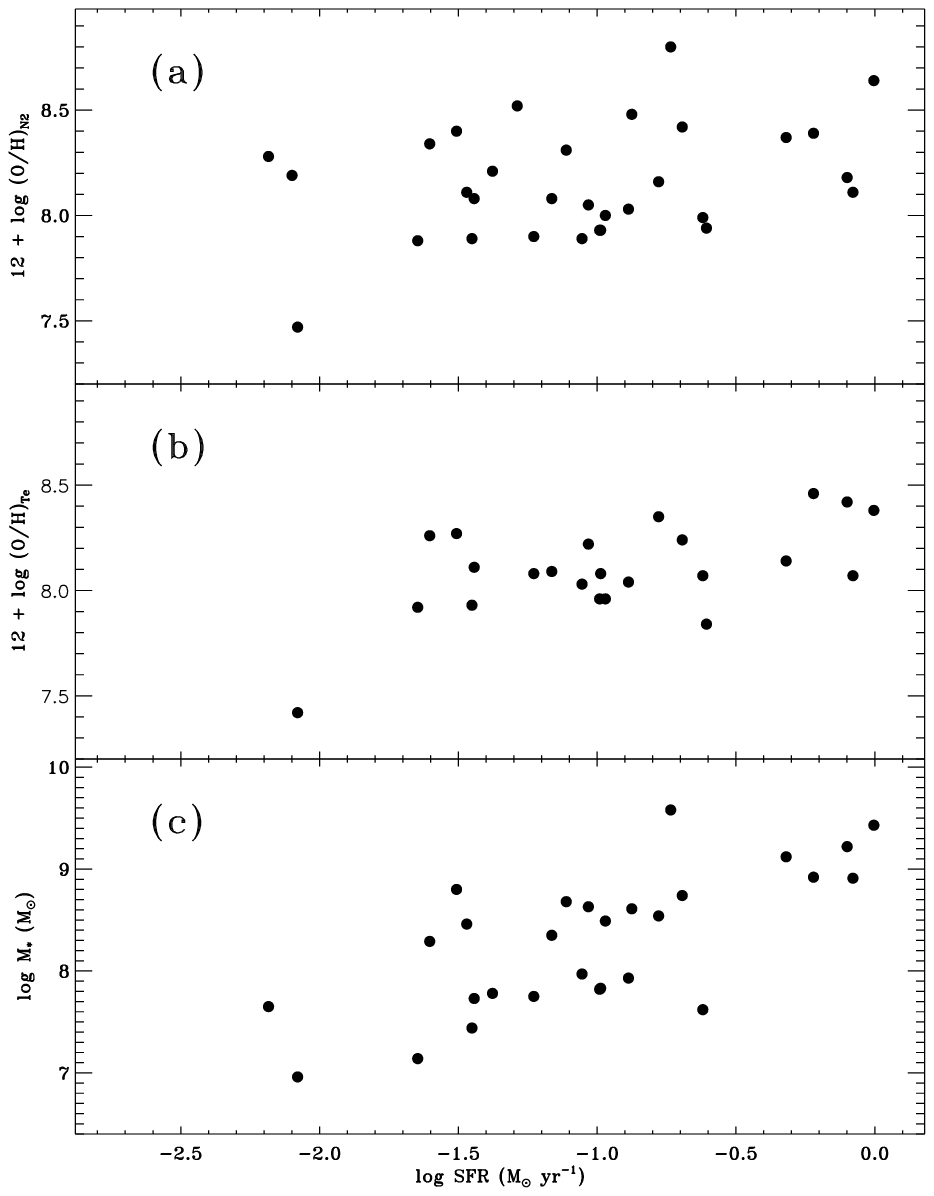}}
\caption{ Correlations between H$\alpha$-based SFR and (a) oxygen abundance derived with the N2-method, (b) oxygen abundance derived with the $T_e$-method, and (c) stellar mass. There are obvious trends for these BCDs with higher stellar mass to also show higher SFRs. However, no clear correlation between metallicity and SFR can be found.}
\label{Fig8}
\end{figure}

Schulte-Ladbeck et al. (2001b) claimed that there might exist a correlation between the oxygen abundance and SFR (see also Hopkins et al. 2002). However, as shown in Figures 8$a$ and 8$b$, we find no evidence of obvious correlation between these two parameters. A Spearman rank correlation coefficient analysis indicates a very weak correlation ($\rho=0.25\ {\rm and}\ 0.38$ at a 1.4$\sigma$ and 1.8$\sigma$ level of significance for Figures 8$a$ and 8$b$, respectively). Therefore, there is little correlation, if any, between SFR and oxygen abundance, which is supported by the fact that there exist large scatters in the plots in Schulte-Ladbeck et al. (2001b; Figure 4) and Hopkins et al. (2002; Figure 6). This argument may also be supported by the results obtained by Kobulnicky \& Skillman (1996, 1997), in which the authors show that their measurements of individual \HII\ regions in dwarf Irregular galaxies suggest localized chemical enrichment from the active star formation generally does not occur.

We can gain further insights on the nature of star formation for BCDs by examining the correlation between galaxy stellar mass and SFR. Noeske et al. (2007), Elbaz et al. (2007), Daddi et al. (2007) and Pannella et al. (2009) have shown that star formation and stellar mass define a tight correlation in galaxies at $z \sim 1$ and $z \sim 2$, with rough proportionality (logarithmic slope of $\sim1.0$). Similar proportionality is also seen at $z = 0$ in data from the SDSS (Elbaz et al. 2007).

We illustrate the relationship between SFR and stellar mass, for our BCDs in Figure 8$c$, which indicates a good correlation between SFR and  M$_{\star}$. A Spearman rank correlation coefficient analysis gives $\rho=0.68$ at a 3.5$\sigma$ level. A nonweighted least-squares linear fit, using a geometrical mean functional relationship (Isobe et al. 1990), to these 28 BCDs gives ${\log\,(\rm{SFR})}\propto (1.21\pm 0.18) \log\,M_{\star}$, which indicates that the SFR increases with stellar mass almost linearly. Our slope of the log(SFR)-log($M_{\star}$) is in agreement with those found by Noeske et al. (2007), Elbaz et al. (2007), Daddi et al. (2007) and Pannella et al. (2009) for more massive local and high-redshift galaxies. However, there is a significant scatter in the trend, i.e., for a given stellar mass, the SFR may vary over an order of magnitude.

\section{Discussion} 
In this section we mainly discuss the uncertainties of the spectral synthesis and the aperture effects.
\subsection{Uncertainties of the Spectral Synthesis}
The SEAGal Group has used the {\sc starlight} to analyze several large samples of SDSS galaxies as shown in their series of works (C05; Cid Fernandes et al. 2007; Mateus et al. 2006; Asari et al. 2007). As mentioned in Section 3.1, they have tested the uncertainties of the resulted stellar populations. In the study of C05, they found that the individual components of {\boldmath $x$} are very uncertain, whereas the binned vectors of {\boldmath $x$}, i.e. the young, inter-mediate and old populations, have uncertainties less than 0.05, 0.1 and 0.1, respectively, for S/N$\geq$10.

For each fitting, {\sc starlight} provides the last-chain-values of the contributed light (mass) fraction of each SSP, $\chi^2$ and mass, for seven Markov chains.  For the results of our sample, the median values of the rms of these adopted values are 1.8\%, 17.4\%, 8.8\%, 8.1\%, 21.0\%, 1.9\%, 0.3\%, 0.5\%, 1.8\%, 0.06\%, 5.1\% for $x_{\rm Y}$, $x_{\rm I}$, $x_{\rm O}$, $\mu_{\rm Y}$, $\mu_{\rm I}$, $\mu_{\rm O}$, $\left<\log t_\star\right>_L$, $\left<\log t_\star\right>_M$, $A^\star_V$, $\chi^2$ and $M_\star$, respectively. Therefore, the uncertainties of the resulting stellar populations and stellar mass will not much affect our conclusions.

Most of the age sensitivity comes from the 4000 \AA\ break and continuum shape in the fitting, thus degeneracy with extinction should be discussed carefully. Stellar mass-related parameters (e.g. $\left<\log t_\star\right>_M$ and $M_\star$) will suffer more severely from this degeneracy because of the non-constant mass-to-light ratio of stars. As described in Section 3.3, the dust extinction is considered as a variable parameter in {\sc starlight}, and it is overestimated (comparing to $A_{V,\,{\rm neb}}$) for at least half of our sample, which might result in a large underestimation of $\left<\log t_\star\right>_M$. For example, out of the 11 galaxies whose fitted $\left<\log t_\star\right>_M$ are less than 1 Gyr based on the fiber spectra (see Table 2), seven have their $A^\star_V$ overestimated. To check to what extent this overestimation will affect the fitted results, we re-run {\sc starlight} using the same parameters except that we limit $A^\star_V$ to the value of $A_{V,\,{\rm{neb}}}$ for ten galaxies whose $A^\star_V/A_{V,\,{\rm neb}}>1.5$. We find that $\Delta\left<\log t_\star\right>_L=0.24\pm0.10$, $\Delta\left<\log t_\star\right>_M=1.15\pm1.03$, and $\Delta \log M_\star=0.41\pm0.27$. Among these ten galaxies, six have $\left<\log t_\star\right>_M < 1$ Gyr (as listed in Table 2), and we obtain $\Delta\left<\log t_\star\right>_L=0.25\pm0.12$, $\Delta\left<\log t_\star\right>_M=1.74\pm0.92$, and $\Delta \log M_\star=0.56\pm0.24$. Therefore, the uncertainties of our fitting results are dominated by the age-extinction degeneracy. 

\subsection{Aperture effects}
The SDSS is a fiber-based survey, and thus we should consider the biases introduced by the use of small apertures to measure the galaxy spectra. Kewley et al. (2005) have examined this problem in detail and they conclude that 20\% of the galaxy light is required to minimize the aperture effects. 

\begin{figure}[pthb]
\centering
\includegraphics[width=0.8\textwidth]{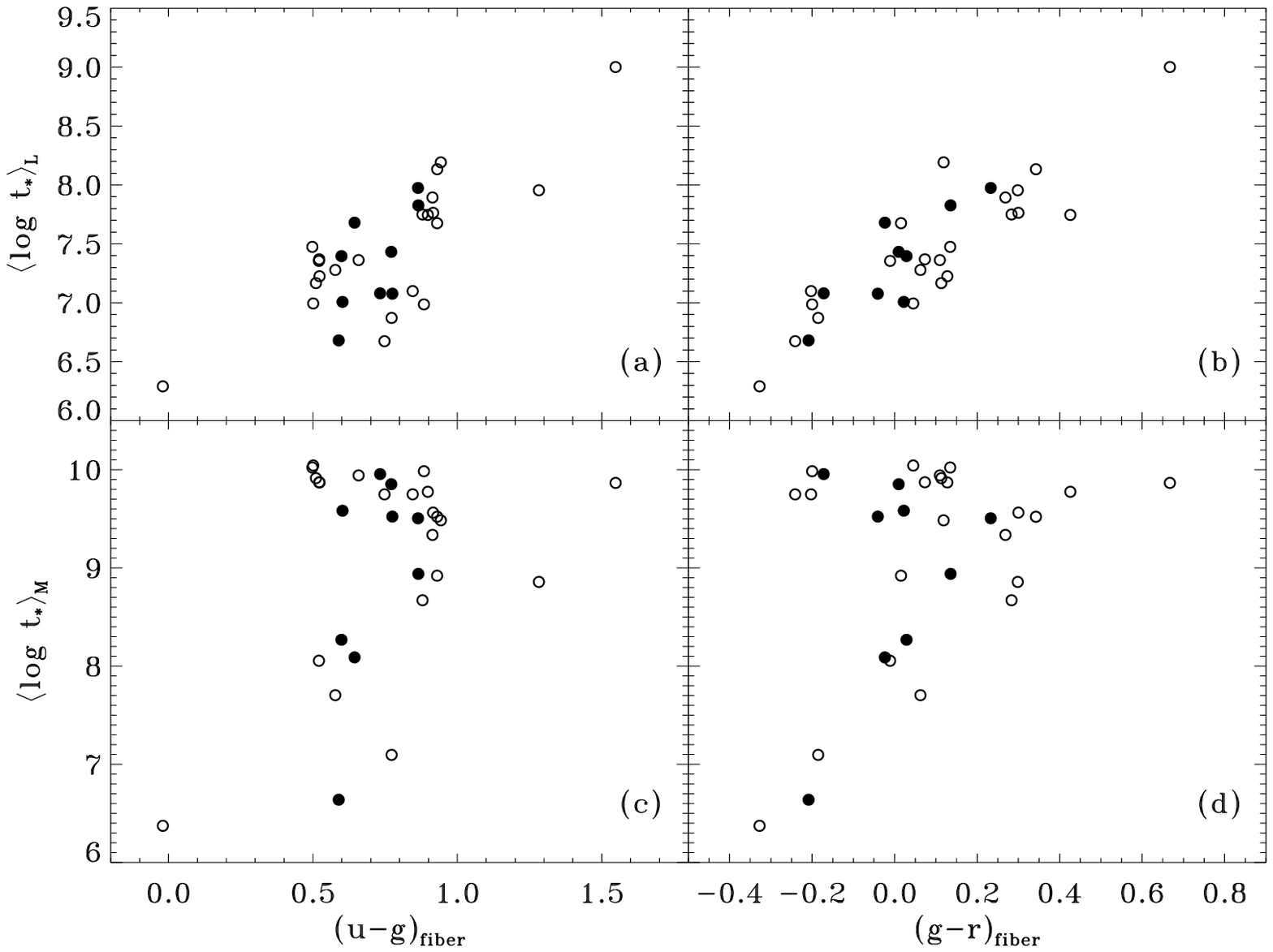}
\caption{ The mean light- (upper panels) and mass-weighted (bottom panels) stellar age as a function of the $(u-g)$ and $(g-r)$ colors within the $3''$-diameter region. Open and solid circles show the spectra with $f_r < 20\%$ and $f_r > 20\%$, respectively.}
\label{Fig9}
\end{figure}

As described earlier, the SDSS spectra for our sample galaxies only cover a small fraction of the galaxy light, and the median $f_r$ is 14.4\%. The fiber spectra are bluer than the integrated SEDs for most galaxies. Based on the fiber spectra, in general, $x_{\rm Y}$ is overestimated and  $x_{\rm O}$ is underestimated (see Figure 6), resulting in an underestimation of both $\left<\log t_\star\right>_L$ and $\left<\log t_\star\right>_M$. Therefore we want to check to what extent our estimates of the stellar properties may be biased on the basis of the fiber spectra. In Figure 9 we plot the dependence of the mean stellar ages on the fiber $(u-g)$ and $(g-r)$ colors. We find that $\left<\log t_\star\right>_L$ correlates both with $(u-g)$ and $(g-r)$ (see Figures 9$a$ and 9$b$). Using these two correlations, we can make a coarse estimation of the bias in $\left<\log t_\star\right>_L$. For our sample, the $3''$-regions are generally 0.15 and 0.17 mag (median value) bluer than the whole galaxies in $(u-g)$ and $(g-r)$, respectively, which result in 0.30 and 0.42 dex underestimations of $\left<\log t_\star\right>_L$ respectively. However, no dependence of $\left<\log t_\star\right>_M$ on these two colors can be established. This is because that $\left<\log t_\star\right>_M$ has much less direct relation with the observed spectrum and is much more affected by the age-extinction degeneracy than $\left<\log t_\star\right>_L$ due to the non-constant mass-to-light ratio of stars. As shown in Section 3.1, the aperture effects generally cause an underestimation of $\left<\log t_\star\right>_M$ for our sample galaxies, and thus, combining with the results derived with the integrated spectra, they might not much affect our conclusion that BCDs are old galaxies experiencing starbursting activities.

\section{Summary}
In this paper we present the detailed results of our stellar population synthesis for an overlapped sample of 31 BCDs between G03 and SDSS DR6, using both integrated and fiber spectra. Our studies show that BCDs are not young systems experiencing their first star formation activities but old systems with episodic star formation histories, which confirm previous studies. For most galaxies, the stellar mass-weighted ages for the $3''$-regions in BCDs are as old as 10 Gyr while the luminosity-weighted ages could be $\sim$2 to 3 orders of magnitude younger. While for the 10 galaxies having integrated spectra, the stellar mass-weighted ages are all larger than several Gyrs. Old population contributes the majority of stellar mass while young population contributes most of light in the optical. 

The SFR of more than half sample galaxies is less than 0.1 $M_\odot$ yr$^{-1}$, and the median SFR is 0.093 $M_\odot$ yr$^{-1}$. For 28 BCDs which have their stellar mass measured reliably, we calculate the birthrate parameter, $b$, and find that about 90\% galaxies have $b>2-3$, which indicates that these galaxies are undergoing a starburst. The current star formation rate correlates well with the integrated galactic stellar mass. The nebular metallicity shows a very weak correlation with the current SFR. 

\acknowledgements
The authors deeply acknowledge the anonymous referee for her/his careful reading and constructive comments which much improved the paper. Y. Zhao is grateful for the financial supports from the NSF of China (grant No. 10903029). Research for this project is partly supported by NSF of China (Distinguished Young Scholars No. 10425313, and grants 10833006, 10621303, 10878010 \& 10633040), Chinese Academy of Sciences' Hundred Talent Program, and 973 project of the Ministry of Science and Technology of China (grants No. 2007CB815405 \& No. 2007CB815406). We gratefully acknowledge Dr. John Moustakas for providing us the integrated spectra. The \emph{starlight} project is supported by the Brazilian agencies CNPq, CAPES and FAPESP and by the  France-Brazil CAPES/Cofecub program. This research has made use of the NASA/IPAC Extragalactic  Database (NED), which is operated by the Jet Propulsion Laboratory, California Institute of Technology, under contract with the National Aeronautics and Space Administration.  All the authors acknowledge the work of the Sloan Digital Sky Survey (SDSS) team. Funding for the SDSS and SDSS-II has been provided by the Alfred P. Sloan Foundation, the Participating Institutions, the National Science Foundation, the U.S. Department of Energy, the National Aeronautics and Space Administration, the Japanese Monbukagakusho, the Max Planck Society, and the Higher Education Funding Council for England. The SDSS Web Site is http://www.sdss.org/. The SDSS is managed by the Astrophysical Research Consortium for the Participating Institutions. The Participating Institutions are the American Museum of Natural History, Astrophysical Institute Potsdam, University of Basel, University of Cambridge, Case Western Reserve University, University of Chicago, Drexel University, Fermilab, the Institute for Advanced Study, the Japan Participation Group, Johns Hopkins University, the Joint Institute for Nuclear Astrophysics, the Kavli Institute for Particle Astrophysics and Cosmology, the Korean Scientist Group, the Chinese Academy of Sciences (LAMOST), Los Alamos National Laboratory, the Max-Planck Institute for Astronomy (MPIA), the Max-Planck-Institute for Astrophysics (MPA), New Mexico State University, Ohio State University, University of Pittsburgh, University of Portsmouth, Princeton University, the United States Naval Observatory, and the University of Washington.

\end{document}